\documentclass{osa-article}

\journal{osac}
\articletype{Research Article}
\usepackage{mathtools}
\usepackage{bm}
\usepackage{float}
\usepackage{url}
\usepackage{subfiles}
\newcommand{\ls}{\left[}
\newcommand{\rs}{\right]}
\newcommand{\lv}{\left|}
\newcommand{\rv}{\right|}
\newcommand{\lp}{\left(}
\newcommand{\rp}{\right)}

\begin{document}

% \title{One-Time Training That Transfers to Arbitrary Highly Faulty Optical Neural Networks}
\title{Transferable Learning on Analog Hardware}

\author{Sri Krishna Vadlamani\authormark{1,*}, Dirk Englund\authormark{1}, and Ryan Hamerly\authormark{1,2}}

\address{\authormark{1}Research Laboratory of Electronics, Massachusetts Institute of Technology, Cambridge, MA, USA\\
\authormark{2}NTT Research Inc., Sunnyvale, CA, USA}

\email{\authormark{*}srikv@mit.edu} %% email address is required



\begin{abstract}
While analog neural network (NN) accelerators promise massive energy and time savings, an important challenge is to make them robust to static fabrication error. Present-day training methods for programmable photonic interferometer circuits, a leading analog NN platform, do not produce networks that perform well in the presence of static hardware errors. Moreover, existing hardware error correction techniques either require individual re-training of every analog NN (which is impractical in an edge setting with millions of devices), place stringent demands on component quality, or introduce hardware overhead. We solve all three problems by introducing one-time error-aware training techniques that produce robust NNs that match the performance of ideal hardware and can be exactly transferred to arbitrary highly faulty photonic NNs with hardware errors up to 5x larger than present-day fabrication tolerances.
\end{abstract}

%%%%%%%%%%%%%%%%%%%%%%%%%%  body  %%%%%%%%%%%%%%%%%%%%%%%%%%
\section{Introduction}

Intense research over the last decade has demonstrated that neural networks possess a remarkable capacity to learn patterns and provide state-of-the-art performance in an astounding variety of artificial intelligence (AI) tasks \cite{silver2018general, brown2020language, ramesh2022hierarchical, jumper2021highly}. Neural networks are parametrized functions that map input vectors to output vectors by performing successive matrix multiplications and elementwise nonlinear operations. The entries of the matrices\textemdash commonly called weights\textemdash are tuned to fit the function model to the training data for the given task. Top-end neural networks today are composed of billions of weights and require massive amounts of data for training. The time and energy costs of training and inference on models of this scale have become a major challenge and have triggered a surge of interest in hardware AI accelerators \cite{9622867, shen2017deep}, both digital and analog. 

\begin{figure}[t]
\centering
\includegraphics[width=\textwidth]{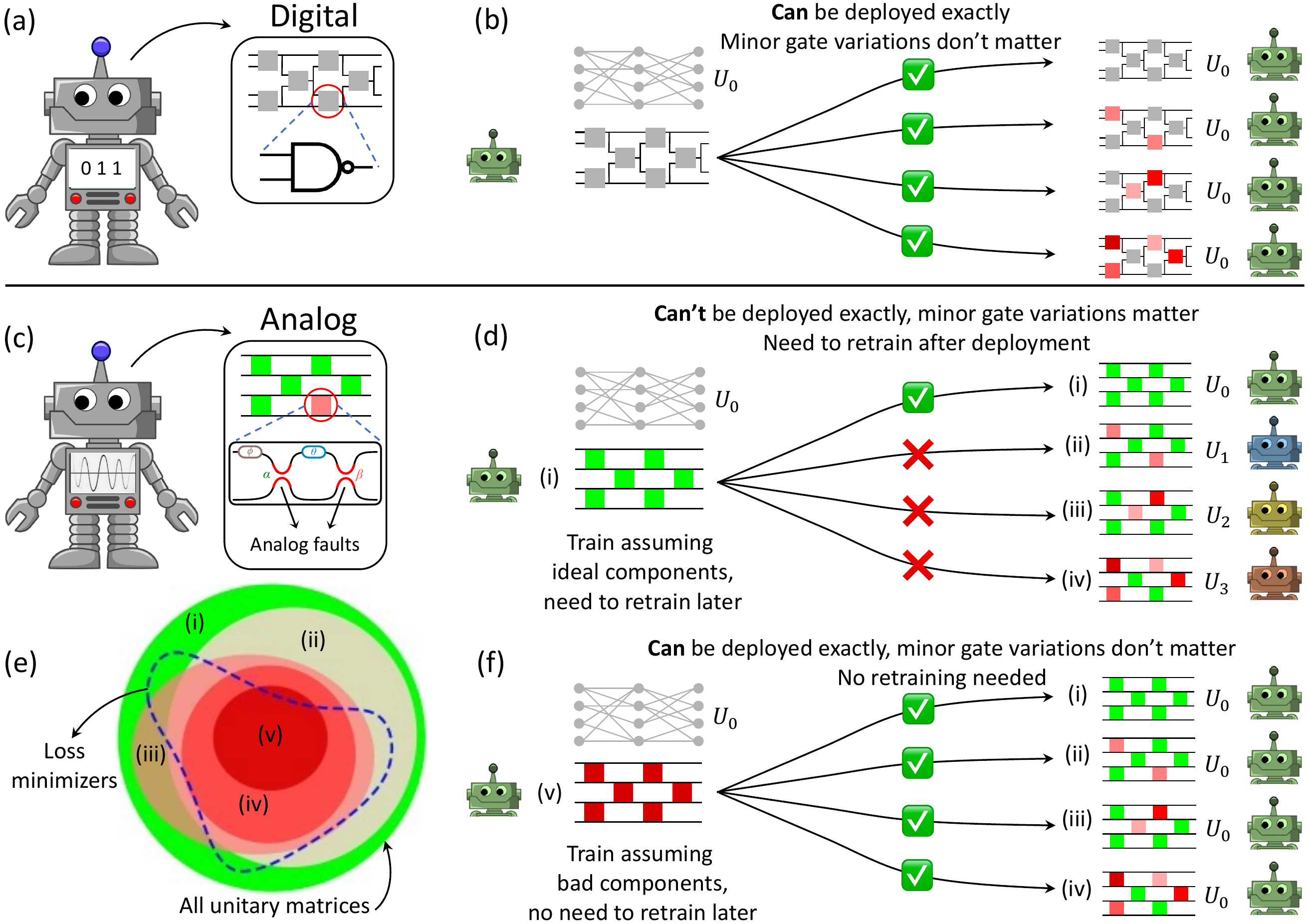}
\caption{(a) Digital AI models run on digital logic circuits. (b) Training is performed once and the resultant model $U_0$ is deployed directly to edge chips which implement $U_0$ faithfully irrespective of their individual gate characteristics. (c) Analog models run on imperfect analog hardware, composed of optical unitary MZI meshes in this case. (d) Training on (a model of) ideal hardware yields function $U_0$ that transfers exactly only to ideal chips. Transferring the raw parameters of $U_0$ to faulty chips leads to undesired functions $U_1,U_2,U_3$ getting implemented. Error-correction or chip-specific retraining do not guarantee recovery of $U_0$. (e) Labels (i)-(v) refer to chips in panel (f). Ideal meshes (chip (i)) can implement all unitary matrices; more faulty meshes implement fewer functions (green shrinks into red). However, there are good loss minimizers (dotted blue line) even within this restricted set. (f) Training on (a model of) very faulty hardware (chip (v)) yields function $U_0$ that transfers exactly to all less faulty chips ((i)-(iv)) with no additional retraining.}
\label{fig:intro}
\end{figure}

Analog accelerators promise tremendous energy and time savings \cite{hamerly2019large, shen2017deep} but one still needs to answer the universal criticism of analog circuits\textemdash that they can be unreliable as general-purpose computers because of both static hardware errors caused by manufacturing variations and inherent noise in the signals being processed. These problems persist in the particular case of analog optical neural networks (ONNs). For instance, the splitting ratio of a typical fabricated beamsplitter deviates by 2\% from 50-50 \cite{mikkelsen2014dimensional} which is sufficient to severely degrade the test accuracy of ONNs \cite{fang2019design} composed of interconnected Mach-Zehnder Interferometers (MZIs) (Fig.~\ref{fig:intro}(c)). Hardware error correction techniques \cite{burgwal2017using, mower2015high, lopez2019programmable, lopez2020auto, perez2020multipurpose, pai2019matrix, hughes2018training, pai2022experimentally, miller2017setting, pai2020parallel, bandyopadhyay2021hardware, hamerly2022design, miller2015perfect, suzuki2015ultra} applied to the hardware parameters provide significant performance improvements but either require individual training/retraining of every ONN \cite{burgwal2017using, mower2015high, lopez2019programmable, lopez2020auto, perez2020multipurpose, pai2019matrix, hughes2018training, pai2022experimentally} (Fig.~\ref{fig:intro}(d)), which is impractical in an edge setting with millions of devices, or place stringent demands on component quality \cite{bandyopadhyay2021hardware}, or introduce hardware overhead \cite{hamerly2022design, miller2015perfect, suzuki2015ultra}. This is in sharp contrast to standard digital NNs (Fig.~\ref{fig:intro}(a)) where training is performed only once and the resultant model can be deployed to any number of devices with no modification (Fig.~\ref{fig:intro}(b)).  

In this paper, we present a one-time error-aware software training technique that solves all three problems at once and brings analog neural networks into the same league as digital neural networks in terms of ease of model training and large-scale deployment. Our method outputs matrices that, while matching the performance of trained ideal hardware, can be exactly transferred to any faulty ONN manufactured by a given process (with no additional training or associated loss of performance, Fig.~\ref{fig:intro}(f)). Moreover, the procedure does not add extra hardware to the existing ONN. We show through numerical simulations that the method in fact tolerates hardware errors up to 5x larger than present-day fabrication errors.

Our method is a combination of two important ideas\textemdash the error-correction scheme of Bandyopadhyay et al. \cite{bandyopadhyay2021hardware}, and a form of engineering corner analysis \cite{orshansky2007design}. In more detail, it is known that splitter faults in an MZI shrink the set of unitary matrices it can implement \cite{hamerly2022design} (Fig.~\ref{fig:intro}(e)); we introduce and train `maximally faulty' MZI mesh-based ONNs that have the most faults and smallest expressivity for a given error level, and show that the resultant matrices both have very high performance and can be exactly transferred to other ONNs with equal or smaller errors. In other words, our one-time training procedure allows us to train only one highly faulty ONN and freely transfer the resulting model to any number of edge ONNs with their own individual faults (Fig.~\ref{fig:intro}(f)). Though we present results for feedforward MZI mesh-based ONNs, the training procedure is applicable to any MZI-based photonic circuit that permits implementation of the error-correction scheme of Ref.~\cite{bandyopadhyay2021hardware}. Our training procedure could, therefore, potentially find use in other applications of photonic circuits \cite{bandyopadhyay2021hardware} such as quantum simulation \cite{harris2017quantum, wang2018multidimensional, qiang2018large, sparrow2018simulating, carolan2015universal}, signal processing \cite{annoni2017unscrambling, ribeiro2016demonstration, milanizadeh2019manipulating, zhuang2015programmable, notaros2017programmable}, and optimization \cite{prabhu2020accelerating}.

%Connecting sentence! raise a big question (and answer it in the affirmative for a special case) Don't bring up Reck and Clements, say any MZI-based scheme. Include Saumil's rings.3-MZI paper, cite the need to get larger bandwidth. Instead of saying present day tolerance, say best .... may do well, we relax the requirements.

The paper is organized as follows: a summary of the optical hardware and the error-correction scheme of Bandyopadhyay et al. \cite{bandyopadhyay2021hardware} is given in Section \ref{structure} in order to make the paper self-contained, maximally faulty MZI meshes are introduced in Section 3, one-time training and numerical results are presented in Section 4, and Section 5 concludes the paper.

\section{Optical neural network structure and error-correction}\label{structure}

\begin{figure}[H]
  \centering
  \includegraphics[width=\textwidth]{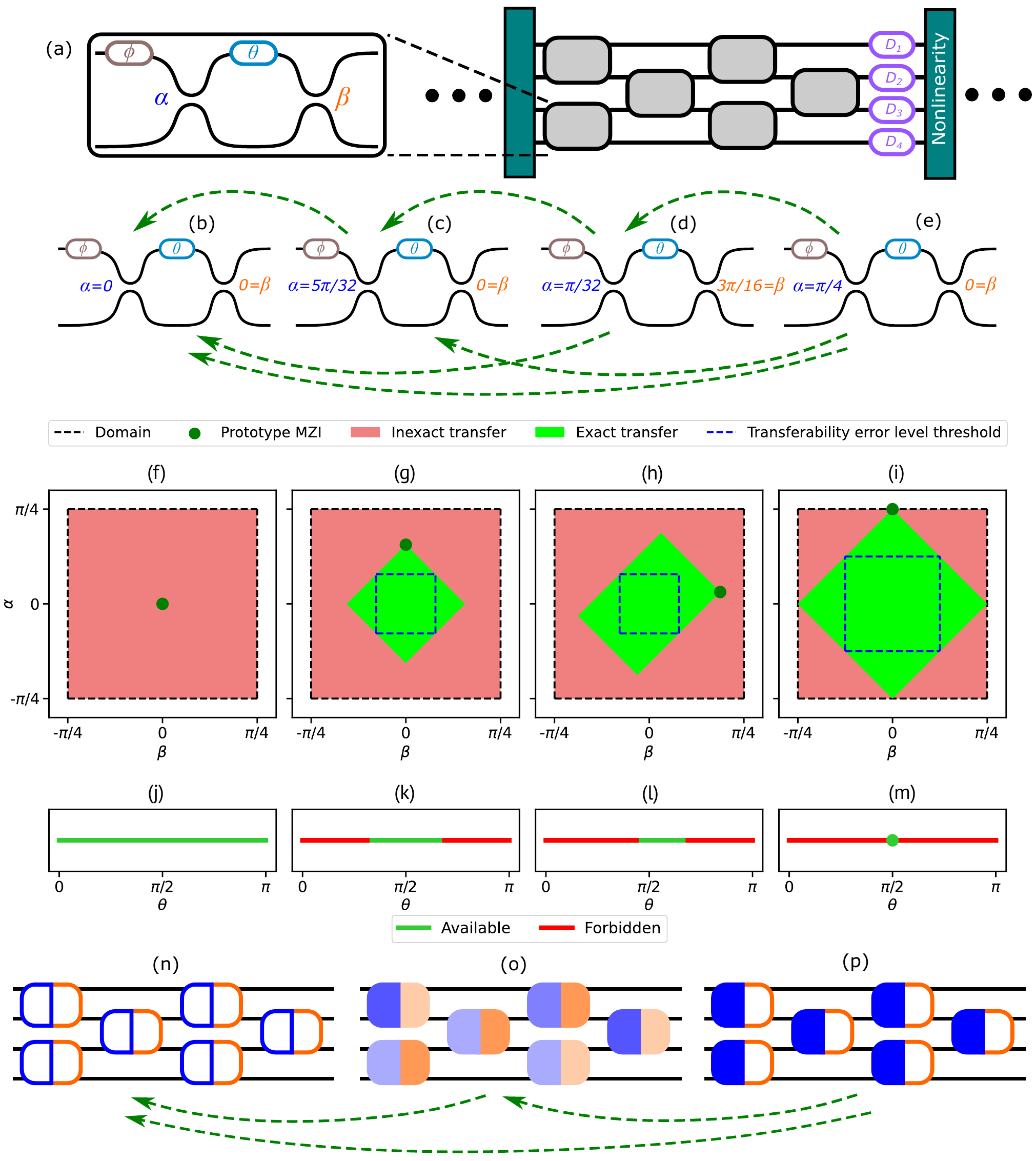}
\caption{(a) 4-by-4 Clements MZI ONN layer. (b) Ideal MZI, (c-e) faulty MZIs. Transfer matrices of more faulty MZIs can be implemented exactly by less faulty MZIs, transferability shown by dashed green arrows. (f-i) Prototype MZI transfers to all MZIs inside the pale green rectangle but not to those outside. The dashed blue square is the largest error level $-\epsilon\leq\alpha,\beta\leq \epsilon$ up to which exact transfer occurs. Prototype MZI errors $(\alpha,\beta)$ of (f-i) are, in units of $\pi$, $(0,0)$, $(5/32,0)$, $(1/32,3/16)$, and $(1/4,0)$ respectively. (j-m) These prototype MZIs implement only a restricted range (green, `Available') of ideal MZI matrix phase-shifts $\theta$. Representing MZI left BS by blue edge color, right by orange, and error level by fill color intensity, (n) ideal mesh, (o) random faulty mesh, (p) maximally faulty mesh with $\alpha=2\epsilon,\beta=0$ everywhere. Matrices of more faulty meshes can be exactly transferred to less faulty ones, transferability shown by dashed green arrows.}\label{fig2}
\end{figure}

Any $N\times N$ unitary matrix can be decomposed \cite{reck1994experimental, clements2016optimal} into a product of $2\times 2$ unitary matrices and an $N\times N$ diagonal matrix $D$ of complex phase-shifts. The $2\times2$ unitaries are implemented in hardware by MZIs while separate phase shifters implement the diagonal matrix (see the circuit between the two nonlinear blocks in Fig.~\ref{fig2}(a)). Each MZI has two phase-shifters, $\theta$ and $\phi$. Individual MZIs are connected in a mesh topology that is consistent with the chosen $N\times N$ unitary decomposition method. ONNs are constructed by interleaving individual $N\times N$ meshes with elementwise nonlinear operations $\sigma(\cdot)$. The nonlinear function implemented by such a network \cite{bandyopadhyay2021hardware} is derived in Supplemental Section 1. Fig.~\ref{fig2}(a) depicts an ONN layer composed of a $4\times4$ rectangular Clements \cite{clements2016optimal} mesh of MZIs. 
 
One way to use ONNs is to train a digital model of an ideal ONN with perfect 50-50 beamsplitters and program the resultant optimal phase-shifts\textemdash $\theta,\phi$ of all the MZIs and the diagonal matrices $D$\textemdash into the hardware for inference. However, as mentioned previously, beamsplitter errors arising from process variation cause a mismatch between the digital model and the model implemented by the hardware, leading to severe degradation of ONN test-time performance when ideal trained phase-shifts are programmed into the faulty hardware with no modification \cite{fang2019design}. Error-correction of some form is therefore essential. 

Published error-correction procedures include global methods that adjust individual MZI phase-angles using circuit-wide optimization \cite{burgwal2017using, mower2015high, lopez2019programmable, lopez2020auto, perez2020multipurpose, pai2019matrix, hughes2018training}, local methods \cite{miller2017setting, pai2020parallel, bandyopadhyay2021hardware} that do so using only device-level information, and hardware augmentation methods which introduce additional beamsplitters (`3-MZIs') \cite{hamerly2022design} or both beamsplitters and phase-shifters \cite{miller2015perfect, suzuki2015ultra} into the system. Global methods can improve performance but can be impractical in edge computing settings where the same model needs to be operated on a large number of edge devices. Local methods apply readily to edge settings because they involve quick local adjustments but they do not correct over a large splitting error range. Hardware augmentation methods correct over a very large error range but incur chip area costs due to the extra hardware. We present a one-time global training method in this paper that readily applies to edge settings, has a large splitting error correction range, and involves no additional hardware overhead. Since our approach uses concepts derived in the local error-correction scheme of Ref.~\cite{bandyopadhyay2021hardware}, we provide a brief overview of their method next.

The transfer function of an imperfect beamsplitter is
\begin{equation}
    T^{\text{bs}}=\begin{pmatrix}\cos{\lp\pi/4+\alpha\rp} &i\sin{\lp\pi/4+\alpha\rp}\\ i\sin{\lp\pi/4+\alpha\rp} &\cos{\lp\pi/4+\alpha\rp}\end{pmatrix}\label{faultytbs}
\end{equation}
where $\alpha$ is the `error angle' that captures the deviation from the ideal 50-50 ratio. Eq.~\eqref{faultytbs} reduces to the 50-50 case for $\alpha=0$. Let the two error angles of a faulty MZI be denoted by $\alpha$ and $\beta$ respectively. Further, let $T(\theta,\phi)$ and $T'(\theta,\phi,\alpha,\beta)$ represent the transfer functions of an ideal and a faulty MZI respectively.

Then, Ref.~\cite{bandyopadhyay2021hardware} shows that one can find an `error-corrected' set of phase-shifts $\theta',\phi',\psi_{1},\psi_{2}$ such that: 
\begin{equation}
    T\lp\theta,\phi\rp=\begin{pmatrix}e^{i\psi_{1}} &0\\0 &e^{i\psi_{2}}\end{pmatrix}\cdot T'\lp\theta',\phi',\alpha,\beta\rp\label{equimatrix}
\end{equation}
if and only if the ideal phase-shift $\theta$ satisfies the following `error-correction condition'
\begin{equation}
    2\lv\alpha+\beta\rv\leq\theta\leq\pi-2\lv\alpha-\beta\rv\label{thetarange}
\end{equation}
In words, Eq.~\eqref{equimatrix} says that an ideal MZI with phase-shifts $\theta$ and $\phi$ can be implemented by programming $\theta'$ and $\phi'$ into an imperfect MZI and adding phase-shifts $\psi_{1}$ and $\psi_{2}$ to the two output arms if and only if the imperfect MZI satisfies the error-correction condition Eq.~\eqref{thetarange}. The phase-shifts $\theta'$ and $\phi'$ can be computed and programmed either via explicit mesh calibration or through self-configuration \cite{hamerly2022accurate, hamerly2022stability, miller2013self, miller2013self2, miller2017setting, grillanda2014non} (Supplemental Sections 3 and 5 contain further details on error-correction). Eq.~\eqref{thetarange} plays a crucial role in the definition of `maximally faulty' MZI meshes, the subject of the next section.

% general figure, inputs into analog circuit, so far faulty analog has failed, here's an example where it works! put it in the figure. intro figure 
  
\section{Maximally faulty MZI meshes}
One approach to error-aware training is to calibrate the errors of the faulty hardware, construct a digital model of the system with the errors taken into account, train the digital model on the given task, and then port the resultant phase-shifts back into the hardware. Alternatively, one could use the `physics-aware training' of Ref.~\cite{wright2022deep} which eliminates the need for explicit error calibration by collecting a training set of input-output pairs of the hardware and training a digital neural network (`digital twin') on it; the digital twin and physical hardware are together used to obtain a good model for the given task. The in-situ training of Refs.~\cite{hughes2018training, pai2022experimentally, bandyopadhyay2022single}, where backpropagation is performed in the hardware itself to obtain mesh-specific matrices, is yet another approach. All these approaches involve training each physical chip with its own individual errors separately (Fig.~\ref{fig:intro}(d)) which is impractical in an edge computing setting with millions of edge devices. To solve this problem, we draw inspiration from corner analysis \cite{orshansky2007design} and introduce the concept of \emph{maximally faulty meshes}\textemdash this idea enables us to train \emph{only one} special mesh (`one-time training') for a given error level $\epsilon$ and transfer the resultant matrices over exactly to any other mesh (having the same geometry) with errors less than $\epsilon$ without any additional mesh-specific training or loss in performance.

More precisely, let us say a fabrication process is guaranteed to produce MZIs with errors $-\epsilon\leq\alpha,\beta\leq \epsilon$ for some error level $\epsilon\geq0$. An MZI is `maximally faulty' for error level $\epsilon$ if its errors satisfy $\alpha=2\epsilon,\beta=0$. A mesh is maximally faulty if all its MZIs are maximally faulty. 

% \begin{figure}[t]
% \centering
% \includegraphics[width=\textwidth]{maxfaultyalphabeta2.pdf}
% \caption{(a-d) Weights of the dark green prototype MZI transfer, through error-correction, to all MZIs inside the pale green rectangle but not to those in the red region. The largest error level $-\epsilon\leq\alpha,\beta\leq \epsilon$ to which weights transfer is given by the largest dashed square (blue) that fits inside the green region. The prototype MZI errors $(\alpha,\beta)$ of panels (a-d) are, in units of $\pi$, $(-1/40,1/20)$, $(1/32,3/16)$, $(5/32,0)$, and $(1/4,0)$ respectively. (e-h) The phase-shifts $\theta$ of ideal MZI matrices that can be implemented exactly by each prototype MZI of the top row are colored green (`Available') in the corresponding panel in the bottom row. The other phase-shifts are in red (`Forbidden').}
% \label{fig:maximallyfaulty}
% \end{figure}

\subsection{Understanding maximally faulty meshes}
To understand the utility of maximally faulty meshes, we return to the error-correction condition Eq.~\eqref{thetarange}. The derivation of the condition implies that any faulty MZI with errors $\alpha,\beta$ can be emulated by an ideal MZI with a $\theta$ that satisfies Eq.~\eqref{thetarange}. The transfer function of this ideal MZI can, in turn, be exactly implemented by any other faulty MZI whose errors $\alpha', \beta'$ satisfy:
\begin{equation}
|\alpha'+\beta'|\leq|\alpha+\beta| \text{ and } |\alpha'-\beta'|\leq|\alpha-\beta|\label{smallerfault}
\end{equation}
since this condition, together with the true statement $2|\alpha+\beta|\leq\theta\leq\pi-2|\alpha-\beta|$, automatically implies $2|\alpha'+\beta'|\leq\theta\leq\pi-2|\alpha'-\beta'|$. In particular, a trained maximally faulty MZI at error level $\epsilon$ can be emulated by any faulty MZI whose errors $\alpha,\beta$ satisfy $|\alpha+\beta|\leq2\epsilon$ and $|\alpha-\beta|\leq2\epsilon$. This set includes all faulty MZIs of error level $\epsilon$, that is, all MZIs in the square bounded by the vertices $(\pm\epsilon,\pm\epsilon)$. We shall refer to the MZI with errors $\alpha,\beta$ as a `prototype' for all other MZIs whose errors $\alpha',\beta'$ are smaller in the sense of Eq.~\eqref{smallerfault}.

Fig.~\ref{fig2}(b)-(e) depict four example MZIs that will be treated as protoype MZIs in this discussion. Fig.~\ref{fig2}(f)-(i) show that prototype MZIs (dark green dot) can be emulated exactly by all MZIs in the $\alpha,\beta$ error phase space that satisfy Eq.~\eqref{smallerfault} (`region of transferability', pale green rectangle) but not by MZIs that don't (red). The errors are specifically chosen such that the prototype MZI of each panel lies within the region of transferability of the prototype MZI of each panel to its right. Therefore, the transfer matrix of panel (e) can be exactly implemented by all the MZIs to its left; the transferability of matrices between meshes is indicated by dashed green arrows. Only the prototype MZIs of panels (c) and (e) are maximally faulty. Fig.~\ref{fig2}(j)-(m) depict the range of ideal MZI $\theta$ phase-shifts that are implementable by the prototype MZIs (b)-(e)\textemdash the more faulty an MZI is, the less expressive it is.

The dashed blue squares (with corners $(\pm \epsilon, \pm \epsilon)$) inside the green rectangles of Fig.~\ref{fig2}(f)-(i) mark the largest error level $\epsilon$ (`transferability error level threshold') up to which the prototype MZI of that panel is transferable. Panels (g) and (h) have dashed blue squares of the same size. However, the maximally faulty MZI of panel (c) explores a wider range of ideal $\theta$ phase-shifts (panel (k)) than MZI (d) (phase-shifts in panel (l)). For any given error level $\epsilon$, maximally faulty MZIs apply less restrictions on the search space $\theta$ than any other prototype MZI.

The transfer of phase-shifts from a more faulty mesh to a less faulty one may be performed in two steps: (1) Translate the phase-shifts of the more faulty mesh to an ideal mesh using the `reverse' of the error-correction of Ref.~\cite{bandyopadhyay2021hardware} (see Supplemental Section 4), (2) translate the ideal phase-shifts to the less faulty mesh using `vanilla' error-correction \cite{bandyopadhyay2021hardware}. Both steps are guaranteed to work exactly. Alternatively, one could use the self-configuration of Refs.~\cite{hamerly2022accurate, hamerly2022stability} to directly program the matrix of the more faulty mesh into the less faulty one in a single step. The all-important role played by the error-correction condition Eq.~\eqref{thetarange} in the above discussion implies that maximally faulty meshes can only be constructed if the underlying mesh geometry permits implementation of the error-correction scheme of Ref.~\cite{bandyopadhyay2021hardware}.

% In summary, it suffices to train a single maximally faulty mesh only once and the resultant matrices can programmed into all chips of error level $y$ with no further chip-specific training or associated loss of performance. Further, the errors of the chips need not even be characterized if the programming is performed via self-configuration \cite{hamerly2022accurate} (also see Supplemental Section 5 for a brief discussion).

\section{One-time training}\label{onetimetraining}
One-time training simply consists of training maximally faulty meshes for a given error level; the resulting matrices can then be transferred directly to any other arbitrary mesh at that error level. We present two approaches to the training: a direct training approach where maximally faulty meshes are trained separately for each given error level, and a transfer training approach where the trained phase-shifts at one error level are used as the starting point for training at the next higher error level. Our simulations (Fig.~\ref{fig:allaccs}(b)) used the \texttt{neurophox} \cite{pai2020parallel} and \texttt{meshes} \cite{hamerly2021meshes} packages, and were performed on an NVIDIA Tesla K40 GPU and the Engaging computing cluster at MIT. Results are presented for the MNIST \cite{lecun2010mnist} (digit), FashionMNIST \cite{DBLP:journals/corr/abs-1708-07747} (clothing), and KMNIST \cite{clanuwat2018deep} (Japanese character) classification tasks.

\subsection{Hyperparameters, pre-processing, baselines}
\begin{figure}[t]
\centering
\includegraphics[width=\textwidth]{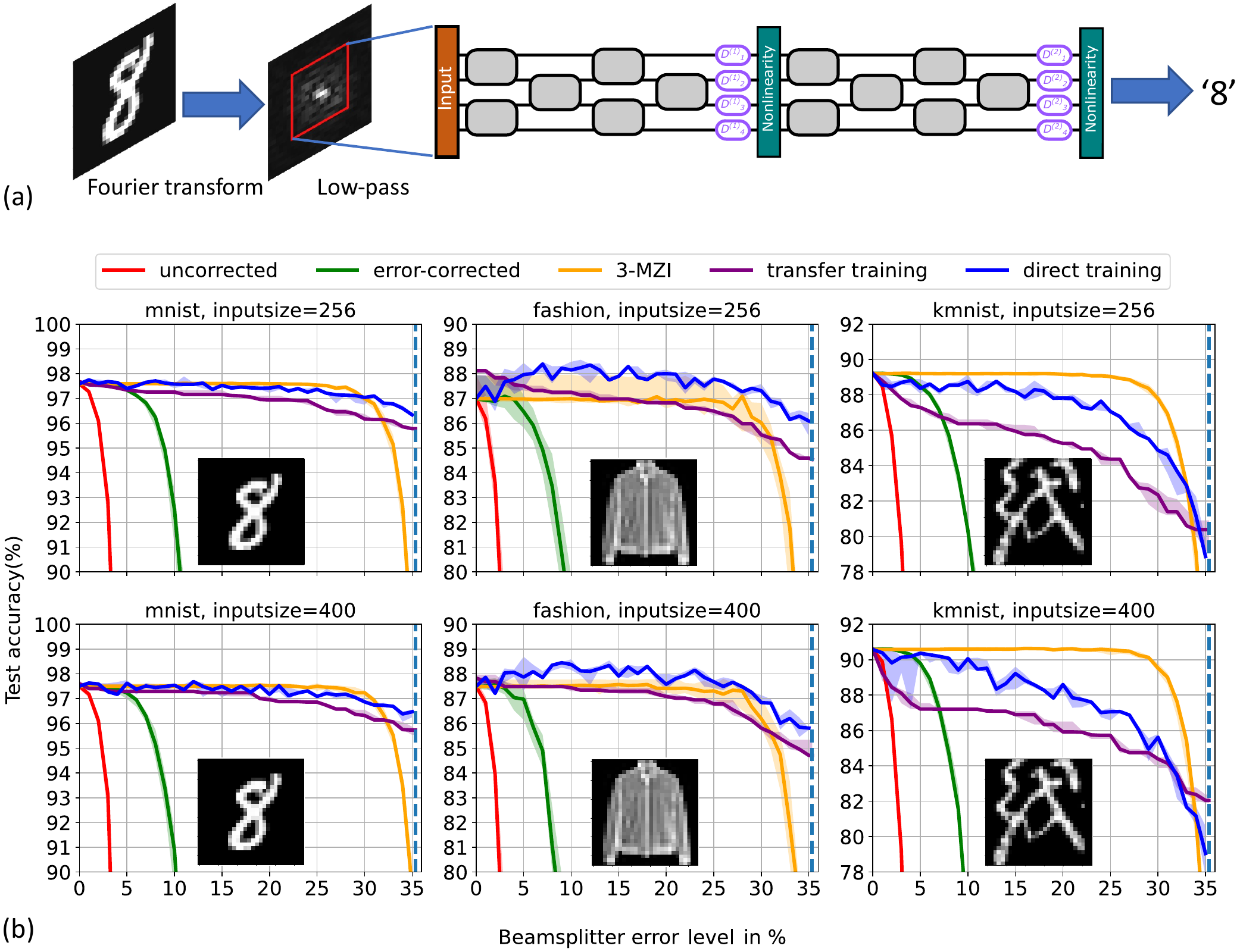}
\caption{(a) Input images are low-pass filtered and passed into a 2-layer ONN of the appropriate size\textemdash layers are depicted with meshes of inputsize 4 for convenience. (b) Performance (median, 25, 75 percentiles) of one-time training on the MNIST, FashionMNIST, and KMNIST datasets, for inputsize 256 and 400. Insets show example images only, the accuracies are computed over the full test set. 5 maximally faulty models were trained for each error level percent from $0\%$ to $35\%$ (step of $1\%$); the test accuracies are plotted in blue (`direct training'). For each of the 5 ideal models, 5 standard MZI and 3-MZI networks with random errors were generated at each error level percent. The test accuracies obtained from direct transfer of ideal model weights to the faulty standard MZI networks are plotted in red, results of transfer after error-correction are in green, results of error-corrected transfer to faulty 3-MZI networks are in yellow. Results of repeatedly transferring lower error level network phase-shifts to a higher error level network and retraining for 2 epochs each time are in purple (`transfer training').}
\label{fig:allaccs}
\end{figure}

Each of the 3 datasets contains 70,000 monochrome images of size $28\times28$\textemdash  60,000 training images and 10,000 test images. The images were low pass filtered by Fourier transforming them and selecting only a smaller square of Fourier components of side $s$ centered about the origin \cite{pai2020parallel}. In order to probe the effectiveness of one-time training at different mesh sizes, we ran simulations for both $s=16$ and $s=20$, which correspond to 256 and 400 total input features (labeled `inputsize' in Fig.~\ref{fig:allaccs}(b)) respectively. Our neural networks were 2-layered, each layer was a Clements unitary mesh, and the electro-optic nonlinearity of Ref.~\cite{williamson2019reprogrammable} was used between layers (Fig.~\ref{fig:allaccs}(a)). The first 10 outputs of the output layer were treated as the label predictors. The standard cross-entropy loss and the Adam optimizer were used.

A closer inspection of Eq.~\eqref{thetarange} reveals a natural upper limit on the error levels that our method can tackle. A maximally faulty MZI emulates ideal MZIs whose phase-shift $\theta$ lies in the range $4\epsilon\leq\theta\leq\pi-4\epsilon$. The lower and upper limits of this range coincide at $\epsilon=\pi/8$ and the expressivity of a maximally faulty MZI collapses to a single value of $\theta$. For $\epsilon>\pi/8$, there doesn't exist a single maximally faulty MZI that transfers to all MZIs at that error level. Therefore, maximally faulty MZIs are a meaningful concept only up to error level $\epsilon=\pi/8$ ($35.36\%$); all our results are plotted up to that error level only.

Fig.~\ref{fig:allaccs}(b) depicts three baselines: (1) the uncorrected case (red), where error-free meshes were trained and the resultant phase-shifts were directly programmed, with no error-correction, into faulty meshes, (2) the corrected case (green), where the ideal-trained phase-shifts were first error-corrected according to Ref.~\cite{bandyopadhyay2021hardware} and then fed into faulty meshes, and (3) the 3-MZI case (orange). 3-MZIs are standard MZIs with an additional beamsplitter \cite{hamerly2022design}. The ideal-trained matrices are fed into faulty 3-MZI meshes via self-configuration \cite{hamerly2022accurate}. 

To generate the baseline data, five ideal meshes with independent Haar-random initial conditions were trained for 50 epochs each. Next, for each ideal model and error level $\epsilon$ (which corresponds to $100\frac{\sin{(2\epsilon)}}{2}$ in percent, the quantity plotted on the x-axis of Fig.~\ref{fig:allaccs}(b)), five faulty meshes were generated with MZI error angles chosen uniformly randomly from the range $[-\epsilon,\epsilon]$. The step size in the error level was $1\%$. The ideal matrices were then transferred to these faulty meshes, by the process appropriate for each baseline, and the test accuracies were recorded. This yields 5 values at each error level for each ideal model. Since there are 5 ideal trained meshes, we have 25 test accuracies for each error level from $1\%$ to $35\%$. The medians of these numbers are plotted as bold lines in Fig.~\ref{fig:allaccs}(b) while the interquartile range (IQR) is represented as a paler sheath of the same color around the central bold line.  

\subsection{Direct training of maximally faulty meshes}
In this approach, the phase-shifts of maximally faulty meshes are trained from randomly initialized starting points $\theta,\phi$ for each error level. For each percentage point between $0\%$ and $35\%$ error level, five maximally faulty meshes were trained independently for 50 epochs and the median and IQR of these 5 values are plotted in Fig.~\ref{fig:allaccs}(b) in blue. In both the MNIST and FashionMNIST tasks, maximally faulty mesh training matches or exceeds the performance of error-correction (green) and the 3-MZI mesh (orange) up to 35\% error level for both mesh sizes considered. There is a curious improvement in performance that direct training achieves compared to the 3-MZI mesh on the FashionMNIST task which one could try to attribute to a regularization caused by the fact that faulty meshes implement fewer unitaries than ideal meshes. That this is not a general phenomenon is immediately borne out by the significantly poorer test accuracy of direct training on KMNIST though it is still within 1\% of the 3-MZI performance up to 10\% error level.

\subsection{Transfer training of maximally faulty meshes}
In this approach, a model for $p\%$ error level is trained using the raw, uncorrected phase-shifts of a trained model for $(p-1)\%$ error level as the starting point. Since model training does not begin from a random starting point, fewer epochs are needed to get to a good set of phase-shifts at the higher error level. In our implementation, we started out once again with the 5 ideal trained models that were previously used for the error-correction and 3-MZI results. The uncorrected ideal model phase-shifts are programmed into a maximally faulty mesh at an error level of $1\%$, and this mesh is trained for 2 epochs. The new phase-shifts are then fed directly into a maximally faulty mesh at an error level of $2\%$ and 2 more epochs of training is performed. This training rate of 2 epochs for every percent increase in error level is maintained up to $35\%$ error level whereupon 2 more epochs of training is performed on a final mesh with $35.36\%$ error level. 

\begin{figure}[t]
\centering
\includegraphics[width=\textwidth]{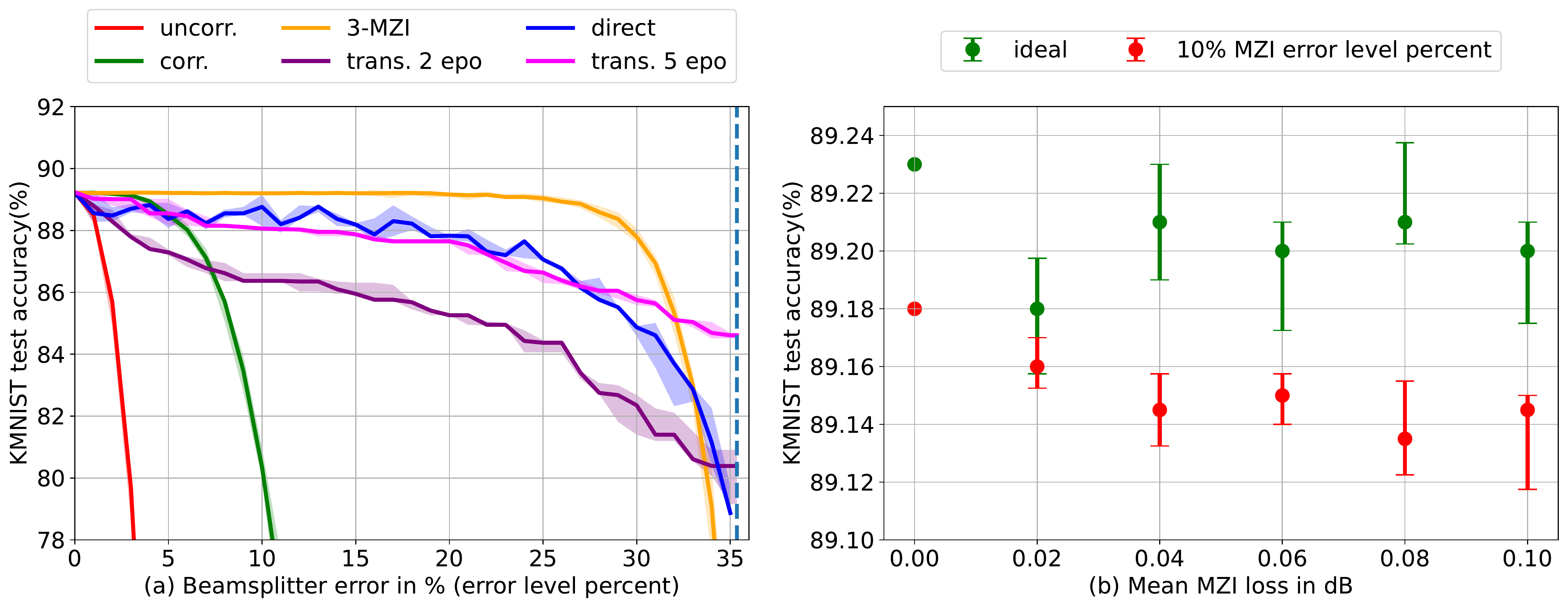}
\caption{(a) Using 5 epochs (pink) of training for every unit increase in error level percent allows transfer training to match direct training for KMNIST, inputsize 256. (b) Direct use of ideal (green) or maximally faulty (red) network weights on the corresponding lossy networks leads to no degradation in test accuracy. 10 lossy meshes were generated at each loss level for both the ideal and the $10\%$ beamsplitter error level percent cases. Results are for KMNIST, inputsize 256.}
\label{fig:trans5epo}
\end{figure}

This procedure is performed with each of the 5 ideal trained models used as a starting point, yielding 5 models at each error level. The test accuracies of these models tend to be non-monotonic, jagged functions of the error level, similar to the jagged blue curves of the direct trained models in Fig.~\ref{fig:allaccs}(b). The fact that the higher error level meshes can be emulated exactly by lower error level meshes suggests that one can make jagged accuracy curves monotonic by assigning to each error level the performance of the best model at the same or higher error level. This approach is computationally prohibitive for direct training since $37\times50=1850$ epochs are required to generate trained models for all error levels from $0\%$ to $35.36\%$. Since it is likely that direct training will be applied to only a few error levels in a real-world setting, Fig.~\ref{fig:allaccs}(b) doesn't depict smoothed out direct training results. On the other hand, since generating transfer trained models for the same range requires only $50+36\times2=122$ epochs, we smooth out the accuracy curves for transfer training and plot the median and IQR in purple in Fig.~\ref{fig:allaccs}(b).

The results indicate that transfer training is nearly as good as direct training and the 3-MZI mesh on both the MNIST and FashionMNIST tasks for both input sizes. On the other hand, the results for KMNIST, which is known to be a difficult dataset \cite{clanuwat2018deep}, are worse than even the error-corrected green curve. Transfer training was re-run for KMNIST, but with 5 epochs of training for every increase in error level compared to the 2 epochs used previously\textemdash the significantly improved performance, which now matches direct training, is depicted in pink in Fig.~\ref{fig:trans5epo}(a). The number of epochs required ($50+36*5=230$) is still smaller than the cost of direct training for all error levels ($1850$ epochs).    

\subsection{Unbalanced MZI losses}
While only splitting errors have been considered so far, unbalanced losses in the MZI arms are another important nonideality. We used the KMNIST inputsize 256 case to demonstrate that unbalanced losses have negligible influence on the performance of maximally faulty meshes. 

MZI loss was assumed to be Gaussian-distributed with the per-MZI mean and variance being integral multiples of $0.02dB$ and $0.0016dB$ respectively \cite{hamerly2022accurate}. Ten random lossy networks with no beamsplitter errors were generated at each loss level, the trained ideal model phase-shifts were programmed directly into them, and the resultant test accuracies are reported in green in Fig.~\ref{fig:trans5epo}(b). The same procedure was also performed for maximally faulty networks with $10\%$ error level; the results are shown in red. It is clear that both types of networks, ideal and maximally faulty, are robust to the unbalanced losses of typical photonic chips.

\section{Conclusion}
Optical neural networks are a leading analog accelerator platform for large-scale machine learning. However, their performance degrades drastically in the presence of static MZI beamsplitter errors \cite{fang2019design}. Existing error-correction procedures are either impractical in large-scale edge settings, applicable over small beamsplitter error ranges, or involve additional hardware overhead. In this paper, we presented a one-time error-aware training technique for MZI-based ONNs that tackles all these problems. The method matches ideal-hardware performance even in the presence of large static hardware phase errors up to 5x larger than present-day fabrication tolerances. Moreover, it is one-time, that is, the training is performed only once and the resultant matrices can be programmed directly into any number of arbitrary highly faulty photonic neural networks in an edge setting. Further, the method uses only standard MZIs and does not require additional hardware.

Our key contribution was the introduction of a principled combination of two important ideas\textemdash error-correction and engineering corner analysis. More specifically, we introduced the concept of a `maximally faulty network'\textemdash one in which every MZI has errors $\alpha=2\epsilon,\beta=0$ for some $\epsilon>0$\textemdash and showed that matrices obtained by training such a network yield excellent test performance over a very large range of $\epsilon$. Furthermore, the trained matrices can be exactly ported, using self-configuration or error-correction, onto other MZI networks (with the same underlying geometry) whose splitting error angles $\alpha,\beta$ all lie in the range $[-\epsilon,\epsilon]$ with no additional training, and no loss of performance, associated with the transfer.

We presented two variants of one-time training: (1) direct training from a randomly initialized starting point $\theta,\phi$ for a given error level and (2) transfer training where one repeatedly transfers trained phase-shifts of networks at lower error levels to networks at a slightly higher error level followed by a small amount of retraining. Numerical experimentation indicated that our method approached or achieved the large error tolerance of 3-MZI \cite{hamerly2022design} networks on several benchmark tasks without incurring the associated additional hardware overhead. We also demonstrated that the typical unbalanced losses of interferometer chips have a negligible effect on the performance of our models, even at a high beamsplitter error level. While our study was based on feedforward MZI ONNs, the procedure can be applied to any MZI-based photonic circuit whose underlying geometry permits implementation of the error-correction scheme of Ref.~\cite{bandyopadhyay2021hardware}. More generally, we believe that the one-time training method is applicable to any physical hardware, even non-optical, that supports some type of error-correction along with a kind of corner analysis which permits exact parameter transfer from a more faulty setup to a less faulty one.

\begin{backmatter}
\bmsection{Funding} S.K.V. was supported by NSF RAISE-TAQS program, grant number 1936314, and the NTT Research Inc. grants "Large-Scale Nanophotonic Circuits for Neuromorphic Computing" and "Netcast", administered by MIT. D.E. acknowledges partial support from programs NSF RAISE-TAQS grant number 1936314 and NSF C-Accel grant number 2040695.

\bmsection{Acknowledgments}
We acknowledge useful discussions with Saumil Bandyopadhyay and Alexander Sludds. The authors thank NVIDIA Corporation for donating the Tesla K40 GPU used in this work. Calculations were also performed on the Engaging computing cluster at MIT. 

\bmsection{Disclosures} The authors are in the process of filing a provisional patent application.

\bmsection{Data availability} Data underlying the results presented in this paper are not publicly available at this time but may be obtained from the authors upon reasonable request.

\bmsection{Supplemental document}
See Supplement 1 for supporting content. 

\end{backmatter}

%%%%%%%%%%%%%%%%%%%%%%% References %%%%%%%%%%%%%%%%%%%%%%%%%

%%%%%%%%%% If using BibTeX:
\bibliography{sample}

\pagebreak
\begin{center}
\textbf{\LARGE Supplemental Material}
\end{center}

\setcounter{section}{0}
\setcounter{equation}{0}
\setcounter{figure}{0}
\setcounter{table}{0}
\setcounter{page}{1}
\renewcommand{\theequation}{S\arabic{equation}}

\vspace{2ex}

\section{Transfer function of an ideal MZI-based neural network}
The formulae in these sections are essentially reproduced from Bandyopadhyay et al. \cite{bandyopadhyay2021hardware}. The transfer function of a perfect 50-50 beamsplitter is:
\begin{equation}
    T^{\text{bs}}=\frac{1}{\sqrt{2}}\begin{pmatrix}1 &i\\ i &1\end{pmatrix}\label{idealtbs}
\end{equation}
The transfer function of the MZI in Fig.~2(a) of the main text, with phase-shifts $\theta,\phi$, is:
\begin{align}
    T\lp\theta,\phi\rp&=\frac{1}{\sqrt{2}}\begin{pmatrix}1 &i\\i &1\end{pmatrix}\cdot\begin{pmatrix}e^{i\theta} &0\\0 &1\end{pmatrix}\cdot\frac{1}{\sqrt{2}}\begin{pmatrix}1 &i\\i &1\end{pmatrix}\cdot\begin{pmatrix}e^{i\phi} &0\\0 &1\end{pmatrix}\\
    &=ie^{i\theta/2}\begin{pmatrix}e^{i\phi}\sin{\lp\theta/2\rp} &\cos{\lp\theta/2\rp}\\e^{i\phi}\cos{\lp\theta/2\rp} &-\sin{\lp\theta/2\rp}\end{pmatrix}\label{idealmatrix}
\end{align}
The total transfer function $T^{(k)}$ of the $k$-th layer in a multilayer feedforward optical neural network is:
\begin{align}
    T^{(k)}&=D^{(k)}\prod_iT\lp\theta^{(k)}_i,\phi^{(k)}_i\rp\label{idealfull}
\end{align} 
where $\theta^{(k)}_i$ and $\phi^{(k)}_i$ are the phase-shifts of the $i$-th MZI in the $k$-th layer and $D^{(k)}$ is the diagonal matrix of the $k$-th layer. Denoting the nonlinear function applied after the $k$-th layer by $\sigma^{(k)}(\cdot)$, the transfer function of a neural network with $L$ layers is:
\begin{equation}
    f^{\text{net}}=\sigma^{(L)}\circ T^{(L)}\circ\sigma^{(L-1)}\circ\dots\circ\sigma^{(1)}\circ T^{(1)}
\end{equation}
where $\circ$ represents function composition. If the $\theta^{(k)}_i$s and $\phi^{(k)}_i$s are collected into vectors $\bm{\Theta}$ and $\bm{\Phi}$ respectively, training the network for a given task involves minimizing the discrepancy (`loss') $L\lp T^{\text{net}}\lp \bm{\Theta}, \bm{\Phi};X\rp, Y\rp$ between the training labels $Y$ and the outputs of the neural network on the training samples $X$.

\section{Transfer function with beamsplitter errors}

The transfer function of an imperfect beamsplitter is:
\begin{equation}
    T'^{\text{bs}}=\begin{pmatrix}\cos{\lp\pi/4+\alpha\rp} &i\sin{\lp\pi/4+\alpha\rp}\\ i\sin{\lp\pi/4+\alpha\rp} &\cos{\lp\pi/4+\alpha\rp}\end{pmatrix}\label{faultytbs}
\end{equation}
where $\alpha$ is an `error angle' that captures the deviation from the ideal 50-50 ratio. The splitting deviation in percent is $100\frac{\sin{\lp2\alpha\rp}}{2}$. In our study, $\alpha$ was limited to the range $\ls-\pi/4,\pi/4\rs$. Eq.~\eqref{faultytbs} reduces to the 50-50 case for $\alpha=0$, while at $\alpha=-\pi/4$ and $\alpha=\pi/4$, the beamsplitter acts as a plain waveguide and a waveguide crossing respectively. Denoting the error angles of the left and right beamsplitters of the MZI by $\alpha$ and $\beta$ respectively, and setting $s:=\alpha+\beta$ and $d:=\alpha-\beta$, the MZI transfer function is:
\begin{align}
    T'\lp\theta,\phi,\alpha,\beta\rp&=ie^{i\theta/2}\begin{pmatrix}\begin{multlined}e^{i\phi}\lp\sin{\lp\frac{\theta}{2}\rp}\cos{\lp d\rp}+\right.\\[-2ex]
    \left.i\cos{\lp\frac{\theta}{2}\rp}\sin{\lp s\rp}\rp
    \end{multlined}
    &\begin{multlined}\lp\cos{\lp\frac{\theta}{2}\rp}\cos{\lp s\rp}+\right.\\[-2ex]
    \left.i\sin{\lp\frac{\theta}{2}\rp}\sin{\lp d\rp}\rp
    \end{multlined}\\[4ex]
    \begin{multlined}e^{i\phi}\lp\cos{\lp\frac{\theta}{2}\rp}\cos{\lp s\rp}-\right.\\[-2ex]
    \left.i\sin{\lp\frac{\theta}{2}\rp}\sin{\lp d\rp}\rp
    \end{multlined}
    &\begin{multlined}\lp-\sin{\lp\frac{\theta}{2}\rp}\cos{\lp d\rp}+\right.\\[-2ex]
    \left.i\cos{\lp\frac{\theta}{2}\rp}\sin{\lp s\rp}\rp\end{multlined}\end{pmatrix}
\end{align}
The transfer function of the $k$-th mesh in a multilayer neural network is: 
\begin{align}
    T'^{(k)}&=D^{(k)}\prod_iT'\lp\theta^{(k)}_i,\phi^{(k)}_i,\alpha^{(k)}_i,\beta^{(k)}_i\rp\label{faultyfull}
\end{align}
where superscript $k$ and subscript $i$ have the same meaning as in the previous section.
% \begin{align}
%     T'\lp\theta'^{(k)}_i,\phi'^{(k)}_i,\alpha^{(k)}_i,\beta^{(k)}_i\rp&=ie^{i\theta'^{(k)}_i/2}\begin{pmatrix}\begin{multlined}e^{i\phi'^{(k)}_i}\lp\sin{\lp\frac{\theta'^{(k)}_i}{2}\rp}\cos{\lp d^{(k)}_i\rp}+\right.\\[-2ex]
%     \left.i\cos{\lp\frac{\theta'^{(k)}_i}{2}\rp}\sin{\lp s^{(k)}_i\rp}\rp
%     \end{multlined}
%     &\begin{multlined}\lp\cos{\lp\frac{\theta'^{(k)}_i}{2}\rp}\cos{\lp s^{(k)}_i\rp}+\right.\\[-2ex]
%     \left.i\sin{\lp\frac{\theta'^{(k)}_i}{2}\rp}\sin{\lp d^{(k)}_i\rp}\rp
%     \end{multlined}\\[4ex]
%     \begin{multlined}e^{i\phi'^{(k)}_i}\lp\cos{\lp\frac{\theta'^{(k)}_i}{2}\rp}\cos{\lp s^{(k)}_i\rp}-\right.\\[-2ex]
%     \left.i\sin{\lp\frac{\theta'^{(k)}_i}{2}\rp}\sin{\lp d^{(k)}_i\rp}\rp
%     \end{multlined}
%     &\begin{multlined}\lp-\sin{\lp\frac{\theta'^{(k)}_i}{2}\rp}\cos{\lp d^{(k)}_i\rp}+\right.\\[-2ex]
%     \left.i\cos{\lp\frac{\theta'^{(k)}_i}{2}\rp}\sin{\lp s^{(k)}_i\rp}\rp\end{multlined}\end{pmatrix}\\
%     T'^{(k),\text{full}}&=D'^{(k)}\prod_iT'\lp\theta'^{(k)}_i,\phi'^{(k)}_i,\alpha^{(k)}_i,\beta^{(k)}_i\rp
% \end{align}
Eqs.~\eqref{idealfull} and \eqref{faultyfull} clearly yield different functions if the same parameters $D^{(k)}$, $\theta^{(k)}_i$, and $\phi^{(k)}_i$ are used in both of them.

\section{Error-correction \textemdash\ converting phase-shifts of an ideal MZI to equivalent phase-shifts for a faulty MZI}
Ref.~\cite{bandyopadhyay2021hardware} showed that an ideal MZI with phase-shifts $\theta,\phi$ can be exactly emulated by a faulty MZI if its errors $\alpha$ and $\beta$ are small enough. More precisely, they showed that one can find an `error-corrected' set of phase-shifts $\theta',\phi',\psi_{1},\psi_{2}$ such that: 
\begin{equation}
    T\lp\theta,\phi\rp=\begin{pmatrix}e^{i\psi_{1}} &0\\0 &e^{i\psi_{2}}\end{pmatrix}\cdot T'\lp\theta',\phi',\alpha,\beta\rp
\end{equation}
In words, this means that the transfer matrix of an ideal MZI with phase-shifts $\theta,\phi$ can be implemented by programming $\theta'$ and $\phi'$ into an imperfect MZI and adding phase-shifts $\psi_{1}$ and $\psi_{2}$ to the two output arms. The MZI in Fig.~1 of the main text, however, doesn't possess phase-shifters on the output arms\textemdash Ref.~\cite{bandyopadhyay2021hardware} solves this by showing that these extra phase-shifts $\psi_1,\psi_2$ can effectively be `pushed' through all the subsequent MZIs and added to the output phase screen $D$. Their error-correction procedure has 4 steps and is discussed next.      

\subsection{Enforcing equality of the ideal and faulty transfer matrix element absolute values}
The first step is to enforce equality of the absolute values of the elements of $T'\lp\theta',\phi',\alpha,\beta\rp$ with the absolute values of the corresponding elements of $T\lp\theta,\phi\rp$. This yields:
\begin{equation}
    \cos{\lp\theta'\rp}=\frac{\cos{\lp\theta\rp}+\sin{\lp2\alpha\rp}\sin{\lp2\beta\rp}}{\cos{\lp2\alpha\rp}\cos{\lp2\beta\rp}}\label{thetaprimeeq}
\end{equation}
This equation has a solution $\theta'$ iff the original angle to be implemented, $\theta$, satisfies: 
\begin{equation}
    -\cos{\lp2\lp\alpha-\beta\rp\rp}\leq\cos{\lp\theta\rp}\leq\cos{\lp2\lp\alpha+\beta\rp\rp}
\end{equation}
which is equivalent to the following if we take $\theta\in[0,\pi]$ and $\alpha,\beta\in\ls-\pi/4,\pi/4\rs$:
\begin{equation}
    2\lv\alpha+\beta\rv\leq\theta\leq\pi-2\lv\alpha-\beta\rv\label{thetarange}
\end{equation}
If $\theta$ lies outside this range, amplitude equality is not satisfied and one cannot implement the ideal transfer matrix using the faulty MZI. Faulty MZIs with smaller splitting angle errors can implement a greater variety of target ideal transfer matrices.  

\subsection{Enforcing equality of the inter-column phase difference of the ideal and faulty transfer matrices}
Next, the phase difference between the columns in both matrices are equalized. If one sets
\begin{align}
    T'_{11}\lp\theta',\phi',\alpha,\beta\rp&=e^{i\phi'}e^{i\zeta'_{11}}\sin{\lp\theta/2\rp}\\
    T'_{12}\lp\theta',\phi',\alpha,\beta\rp&=e^{i\zeta'_{12}}\cos{\lp\theta/2\rp},
\end{align}
their ratio is $e^{i\lp\phi'+\zeta'_{11}-\zeta'_{12}\rp}\tan{\lp\theta/2\rp}$. On the other hand, the ratio $T_{11}\lp\theta,\phi\rp/T_{12}\lp\theta,\phi\rp$ is $e^{i\phi}\tan{\lp\theta/2\rp}$. Enforcing equality between the two yields:
\begin{equation}
    \phi'=\phi-\zeta'_{11}+\zeta'_{12}\label{phiprimeeq}
\end{equation}
Thus $\phi'$ can be readily calculated from $\theta'$ and $\theta$. However, this $\phi'$ is only an intermediate value; step D below involves further changes to $\phi'$.

\subsection{Enforcing equality of the global phases of the rows of the ideal and faulty transfer matrices}
Using the $\theta'$ and $\phi'$ computed above, the transfer function of the faulty MZI is:
\begin{equation}
    T'\lp\theta',\phi',\alpha,\beta\rp=ie^{i\theta'/2}\begin{pmatrix}e^{i\zeta'_{12}} &0\\0 &e^{i\zeta'_{22}}\end{pmatrix}\cdot\begin{pmatrix}e^{i\phi}\sin{\lp\theta/2\rp} &\cos{\lp\theta/2\rp}\\e^{i\phi}\cos{\lp\theta/2\rp} &-\sin{\lp\theta/2\rp}\end{pmatrix}\label{almostidealmatrix}
\end{equation}
To make Eq.~\eqref{almostidealmatrix} match Eq.~\eqref{idealmatrix}, phase-shifts of $\psi_{1}=\lp\theta-\theta'\rp/2-\zeta'_{12}$ and $\psi_{2}=\lp\theta-\theta'\rp/2-\zeta'_{22}$ are needed on the first and second output ports of the faulty MZI respectively.       

\subsection{Sending intermediate output phases to the final output phase screen}
Since the MZI architecture in Fig.~1 of the main text does not allow phase shifts on the output ports, the following identity is used to `pass' the phases through the subsequent MZIs in the mesh:
\begin{equation}
    T'\lp\theta',\phi',\alpha,\beta\rp\cdot\begin{pmatrix}e^{i\chi_1} &0\\0 &e^{i\chi_2}\end{pmatrix}\equiv\begin{pmatrix}e^{i\chi_2} &0\\0 &e^{i\chi_2}\end{pmatrix}\cdot T'\lp\theta',\phi'+\chi_1-\chi_2,\alpha,\beta\rp
\end{equation}
The output phases of all the MZIs in the mesh are passed through to the mesh output and added to the existing output phase screen $D$ to yield a new output phase screen $D'$. This step modifies all the $\phi'$ previously computed in step B. This completes the discussion of the error-correction procedure of Ref.~\cite{bandyopadhyay2021hardware}.

\section{Error-correction \textemdash\ converting phase-shifts of a faulty MZI to the equivalent phase-shifts for an ideal MZI}
The converse problem of finding phase-shifts $\theta,\phi$ for an ideal MZI so that it exactly implements a faulty MZI with phase-shifts $\theta',\phi'$ has a similar 4-step solution.

\subsection{Enforcing equality of the ideal and faulty transfer matrix element absolute values}
The first step is identical. Given a $\theta'$, Eq.~\eqref{thetaprimeeq} is solved for $\theta$. The difference from before is that this direction is guaranteed to always have a solution.

\subsection{Enforcing equality of the inter-column phase difference of the ideal and faulty transfer matrices}
Next, the phase difference between the columns in both matrices are equalized. If one sets
\begin{align}
    T_{11}\lp\theta,\phi\rp&=e^{i\phi}e^{i\zeta_{11}}\lp\sin{\lp\frac{\theta'}{2}\rp}\cos{\lp d\rp}+i\cos{\lp\frac{\theta'}{2}\rp}\sin{\lp s\rp}\rp:=e^{i\phi}e^{i\zeta_{11}}X\\
    T_{12}\lp\theta,\phi\rp&=e^{i\zeta_{12}}\lp\cos{\lp\frac{\theta'}{2}\rp}\cos{\lp s\rp}+i\sin{\lp\frac{\theta'}{2}\rp}\sin{\lp d\rp}\rp:=e^{i\zeta_{12}}Y,
\end{align}
where we used $X$ and $Y$ to represent the terms inside the parentheses, the ratio $T_{11}\lp\theta,\phi\rp/T_{12}\lp\theta,\phi\rp$ is $e^{i\lp\phi+\zeta_{11}-\zeta_{12}\rp}\frac{X}{Y}$. On the other hand, the ratio $T'_{11}\lp\theta',\phi',\alpha,\beta\rp/T'_{12}\lp\theta',\phi',\alpha,\beta\rp$ is $e^{i\phi'}\frac{X}{Y}$. Enforcing equality between the two yields:
\begin{equation}
    \phi=\phi'-\zeta_{11}+\zeta_{12}\label{phiprimeeq}
\end{equation}
Thus $\phi$ can be readily calculated from $\theta'$ and $\theta$. However, this $\phi$ is only an intermediate value; step D will modify $\phi$.

\subsection{Enforcing equality of the global phases of the rows of the ideal and faulty transfer matrices}
Using the $\theta$ and $\phi$ computed above, the transfer function of the ideal MZI is now:
\begin{equation}
    T\lp\theta,\phi\rp=e^{i(\theta-\theta')/2}\begin{pmatrix}e^{i\zeta_{12}} &0\\0 &e^{i\zeta_{22}}\end{pmatrix}\cdot T\lp\theta',\phi',\alpha,\beta\rp\label{almostfaultymatrix}
\end{equation}
To make the right-hand side of Eq.~\eqref{almostfaultymatrix} exactly equal to $T'\lp\theta',\phi',\alpha,\beta\rp$, phase-shifts of $\psi_{1}=\lp\theta'-\theta\rp/2-\zeta_{12}$ and $\psi_{2}=\lp\theta'-\theta\rp/2-\zeta_{22}$ are needed on the first and second output ports of the ideal MZI respectively.       

\subsection{Sending intermediate output phases to the final output phase screen}
The extra output phases are passed through the mesh using the same identity as before:
\begin{equation}
    T\lp\theta,\phi\rp\cdot\begin{pmatrix}e^{i\chi_1} &0\\0 &e^{i\chi_2}\end{pmatrix}\equiv\begin{pmatrix}e^{i\chi_2} &0\\0 &e^{i\chi_2}\end{pmatrix}\cdot T\lp\theta,\phi+\chi_1-\chi_2\rp
\end{equation}
and the existing output phase screen $D'$ gets modified to a new phase screen $D$.

\section{Self-configuration}
While the error-correction algorithm of Ref.~\cite{bandyopadhyay2021hardware} requires knowledge of the MZI errors $\alpha,\beta$ to program a given unitary matrix into a faulty MZI mesh, the self-configuration scheme of Hamerly et al. \cite{hamerly2022accurate} can achieve the same without knowledge of the errors. Ref.~\cite{hamerly2022accurate} sets MZI phase-shifts by sending special inputs into the mesh and tuning the phase-shifts until specific expected outputs are obtained. In particular, the scheme self-configures a mesh architecture by cleverly exploiting the unitary decomposition method that led to that architecture in the first place. 

The standard Reck \cite{reck1994experimental} and Clements \cite{clements2016optimal} unitary decompositions proceed by repeatedly applying Givens rotations on the original unitary matrix until a diagonal matrix is obtained. Each Givens rotation zeros out an off-diagonal element of the unitary matrix it operates on; a total of ${N}\choose{2}$ rotations are required to complete the decomposition. An important point here is that the rotations that appear later in the sequence do not disturb the zeros created by earlier rotations. Ref.~\cite{hamerly2022accurate} self-configures meshes by first decomposing the target matrix using the method appropriate for the mesh architecture. Then, for each MZI, a specific column of a running product of the Givens rotations is sent in as input and the phase-shifts of that MZI are tuned until a specific output is zero. The order in which MZIs are tuned depends on the decomposition method used. Once an MZI is tuned such that a particular output is zero for a specific input, that relation will hold even if MZIs downstream are tuned later on. This follows from the aforementioned fact that later rotations do not affect zeros created by earlier rotations. In effect, Ref.~\cite{hamerly2022accurate} implements unitaries by imitating the `decomposition by zeroing out' process in hardware. Readers are referred to Ref.~\cite{hamerly2022accurate} for further details.

%%%%%%%%%% If preparing manually:
% \newcommand{\enquote}[1]{``#1''}

% \bibitem{Zhang:14}
% Y.~Zhang, S.~Qiao, L.~Sun, Q.~W. Shi, W.~Huang, L.~Li, and Z.~Yang,
%   \enquote{Photoinduced active terahertz metamaterials with nanostructured
%   vanadium dioxide film deposited by sol-gel method,}
%   {\protect\JournalTitle{Optics Express}} \textbf{22}, 11070--11078 (2014).

% {Optical Society}, \enquote{{OSA Publishing},}
%   \url{http://www.osapublishing.org}.

% \bibitem{FORSTER2007}
% P.~Forster, V.~Ramaswamy, P.~Artaxo, T.~Bernsten, R.~Betts, D.~Fahey,
%   J.~Haywood, J.~Lean, D.~Lowe, G.~Myhre, J.~Nganga, R.~Prinn, G.~Raga,
%   M.~Schulz, and R.~V. Dorland, \enquote{Changes in atmospheric consituents and
%   in radiative forcing,} in \enquote{Climate Change 2007: The Physical Science
%   Basis. Contribution of Working Group 1 to the Fourth assesment report of
%   Intergovernmental Panel on Climate Change,}  S.~Solomon, D.~Qin, M.~Manning,
%   Z.~Chen, M.~Marquis, K.~B. Averyt, M.~Tignor, and H.~L. Miler, eds.
%   (Cambridge University Press, 2007).

% \end{thebibliography}

\end{document}